\documentclass[aps,preprint]{revtex4}

\usepackage{graphicx}

\usepackage{natbib}

\newcommand{\be}{\begin{equation}}
\newcommand{\ee}{\end{equation}}
\newcommand{\nn}{\mbox{} \nonumber \\ \mbox{} }
\newcommand{\ba}{\begin{eqnarray}}
\newcommand{\ea}{\end{eqnarray}}
\newcommand{\om}{\omega}

\newcommand\eg{\textit{e.g.}}

\newcommand\ie{\textit{i.e.}}

\newcommand{\Bf}{{magnetic field}}

\newcommand{\NS}{neutron star}
\newcommand{\ms}{magnetosphere}
\newcommand{\LC}{light cylinder}

\newcommand{\Sc}{Schwarzschild}
\newcommand{\BH}{black hole}
\newcommand{\BHs}{black holes}
\newcommand{\cfg}{centrifugal}

\begin{document}

\title{On reversal of centrifugal acceleration in    special relativity }
\author{Maxim Lyutikov}
\address{
Department of Physics, Purdue University, 
 525 Northwestern Avenue,
West Lafayette, IN
47907-2036 }

\begin{abstract}
The basic principles of General Theory of Relativity historically have been tested in gedanken   experiments in rotating frame of references. One of the key issues, which still evokes a lot of controversy, is the centrifugal  acceleration.
Machabeli \& Rogava (1994) argued that centrifugal  acceleration reverse direction for particles  moving radially with  relativistic velocities within a "bead on a wire" approximation. We show  that this result is  frame-dependent and reflects a special relativistic dilution of time (as correctly argued by de Felice (1995)) and is analogous to freezing of motion on the black hole horizon as seen by a remote observer.  It is a reversal of  coordinate acceleration; there is no such effect
as measured by a defined set of observers, \eg,  proper and/or comoving.  Frame-independent velocity of a "bead" with respect to stationary rotating observers increases and  formally reaches the speed of light on the light cylinder. In general relativity, centrifugal  force does reverse its direction at photon circular orbit, $r=3M$ in \Sc\ metric, as argued by Abramowicz (1990). 
\end{abstract}

\maketitle

\section{Introduction}
\label{intro}

Since the conception of the Special and General Theories of Relativity, rotating frames served as conceptual testbed of our understanding of effects of motion  and gravitation on measured quantities. For over a century this has lead to a number of paradoxes, most notably  the Ehrenfest paradox \cite{Ehrenfest}  of the circumference length of a rotating disk. The  Ehrenfest paradox involved a discussion between such prominent physicists as Born, Plank, Kaluza, Einstein, Becquerel, and  Langevin among others \citep{Gron}. Kinematics and especially dynamics in rotating frame continues to be a source of confusion. In this article  we aim to elucidate one of the ``paradoxes'', the reversal of centrifugal acceleration. 

Following the work 
\cite{AbramowiczPrasanna} 
on reversal of \cfg\ force in general relativity, 
 Machabeli \& Rogava  \cite{MachabeliRogava}
 suggested that a somewhat similar effect, reversal of \cfg\ acceleration, occurs in special relativity. This suggestion we taken up in a number of astrophysically-related works on particle acceleration around rotating \BH\ and \NS\ magnetospheres \cite{RiegerMannheim, ChiconeMashhoon,ThomasGangadhara,RiegerAharonian} and others. In this Letter we show that the  effect discussed by  
  Machabeli \& Rogava \cite{MachabeliRogava} is not frame-invariant and disappears if one uses frame-invariant quantities. Thus, in special relativity, there is no reversal of  \cfg\  acceleration. The effect seen by Machabeli \& Rogava is a time dilation, as correctly argued by \cite{deFelice}. It describes an unphysical coordinate acceleration.
  
The motivation for this work comes from  numerous astrophysical cites (\eg,  {\ms}s of various \BHs\ and {\NS}s),  where both strong gravity, \Bf, and rotation are all important ingredients. The effects of \Bf\ on a single
particle motion are often approximated as a solid guiding wire, which restricts particle motion across the field. This simple approximation
 neglecting various cross-field drifts.  The key question that we will address is ``what is the behavior of the parallel momentum of the particle?".

 \section{Rotating wire}

\subsection{Motion in coordinate time}

To elucidate the key problems, 
consider a  bead  on a radial  wire inclined at angle $\pi/2$ to the rotation axis. Let us first neglect gravitation.
Using standard methods of general relativity, we  transform to rotating coordinates by  changing  the azimuthal variable  $\phi \rightarrow \phi' - \om t$ 
 and
assume  that in rotating  coordinates the  motion is strictly radial, $d\phi'=d\theta =0$. The non-trivial element of the metric tensor is then 
\be
g_{00} = - \left( 1  - \om^2 r^2 \right)
\label{g00}
\ee
The Hamilton - Jacobi equation
$\partial _t S + H =0$,
where $H$ is Hamiltonian and $S$ is generating function, then becomes
\be
{1\over  1 - \om^2 r^2} (\partial_t S)^2 -  (\partial_r S)^2 =1
\label{H}
\ee
 (we set $G=c=1$, use $(-1,1,1,1)$ sign convention and assume that the mass of a test particle is unity.)
Since  the two-dimensional motion in $r-t$ plane has a conserved
quantity  --  the product of the particle momentum and the time-like Killing vector (time is  a cyclic variable), we can look for separable solutions in a form 
$S= - E_0 t + S(r)$. After differentiating with respect to $E_0$,  Eq. (\ref{H}) gives
\be
(\partial_t  r)^2 =  \left( 1 - \om^2 r^2 \right) 
 \left( 1 - {1 - \om^2 r^2  \over E_0^2}  \right)
 \label{E0}
 \ee
 By differentiating with respect to time, and eliminating  constant $E_0$, we find expression for 
  coordinate acceleration in terms of coordinate velocity
  \be
\ddot{r}= r \om^2   \left( 1 - { 2 v_r ^2 \over 1 - r^2 \om^2  } \right)
\label{a1}
\ee
where $v_r = \partial r /\partial t$. (This result can also be heuristically obtained from  Newtonian centrifugal acceleration formula 
$\partial_t (m_{eff}\partial _t{r}) = r  \om^2 m_{eff} $ with $m_{eff} =1/\sqrt{1- r^2 \om^2 -\dot{r}^2} $.) This is the result of   
 Machabeli \& Rogava \citep{MachabeliRogava},
 who argued that at $r=0$, 
for $v_r > 1/\sqrt{2}$ \cfg\ acceleration reverses its sign and becomes \cfg\  deceleration. Indeed, for  $v_r > 1/\sqrt{2}$ we have $\ddot{r} <0$. In addition,
Eq. (\ref{a1}) does have a solution in terms of elliptic sinus function with formal reversal of velocity occurring at the \LC. 

\subsection{Coordinate and physical acceleration}

In the previous section we derived equations of motion of a bead on a wire and obtained  fully analytical and mathematically correct solutions.
Does it mean that a particle experiences a reversal of centrifugal accelerations and can never leave the \LC\ of a  rigidly rotating wire?
The answer,  which is physically obvious, but given the above derivation is a bit surprising,   is no.
The key moment missed by Machabeli \& Rogava is that 
{\it  observed quantities must be  formulated in a frame-invariant, but observer-dependent form. }
Thus, quantities measured in terms of, \eg,  coordinate time are, in some sense, the least physical. On the other hand, quantities measured by a defined set of observers can be cast in a frame-independent form using the four-velocities of those observers (\eg, notion of ZAMOs in  \cite{MTW}). 
Expression (\ref{a1}) is  
coordinate-dependent, and thus   is not physically useful. Physically important are velocities and  acceleration  measured by a  defined set of observers.  For example, we can define a set  of 
 local stationary observers rotating with the wire.
 For such  observers
$
  dr_s ={dr  }, \,
 dt_s = \sqrt{1- \om^2 r^2 } dt$, 
 \ba &&
(\partial_{t _s} r_s)^2 = 1- { 1- \om^2 r^2   \over E_0^2} 
\nn &&
 { \partial^ 2r_s \over \partial t_s^2 }=  { r \om^2 \over E_0^2}=
   { r \om^2   (1 - (\partial_{t _s} r_s)^2)  \over 1- \om^2 r^2 }
 \label{AA}
 \ea
 Eqns (\ref{AA}) clearly shows that centrifugal acceleration of the bead, as measured by a set of  observers stationary  with respect to the rotating wire, is always directed away from the axis of rotation.

 We can also find equations of motions and acceleration in terms of proper time $\tau$ of the bead:
 \ba &&
 (\partial_\tau r)^2 = {E_0^2 \over 1- r^2 \Omega^2} -1
 \nn &&
 \partial_\tau^2 r= r \Omega^2 {E_0^2 \over (1- r^2 \Omega^2)^2}>0
 \ea
 So that proper velocity  and proper acceleration are  always positive.

As the question under consideration is  controversial, we next show that  
 that  velocity (\ref{AA}), \ie\  the velocity of a bead with respect to rotating observer staitonary with respect to the wire, is frame-invariant.  Recalling  that
a {\it  frame-independent } value of relative velocity of two observers $V_{rel}$  moving with four-velocities
$U$ and $V$ can be calculated according to
\be
V_{rel}^2 = 1 -{1 \over (U  \cdot V)^2}.
\ee
Using (\ref{g00}), the
velocity of a stationary observer at $r$ in coordinates $\{t, r\}$
is 
\be
U^\mu=\{ - {1\over \sqrt{  1 -r ^2 \om^2}},0\}
\ee
Radial velocity in coordinate time is  (\ref{E0}), so that 
 \be
 V^\mu = \{ - { E_0  \over 1- r ^2 \om^2},  \sqrt{ {E_0^2 \over 1-r ^2 \om^2} -1} \}
\ee
(recall, that it is the use of this expression that leads to "reversal" of \cfg\ acceleration.)
The relative velocity between the bead and local stationary observer is
\be 
V_{rel}^2 = 1 -{1- r ^2 \om^2 \over E_0^2},
\label{V}
\ee
consistent with (\ref{AA}).

Eqns. (\ref{AA})  and (\ref{V}) shows that velocity of the bead as measured by a defined set of observers, \eg\ stationary with respect to the wire, increases toward the \LC\ and becomes $c$.  {\it This is  a frame-independent statement: velocity of a bead measured by any observer reaches $c$ on the light cylinder. }

Eq. (\ref{AA}) can be integrated, assuming that a particle starts with velocity $v_0$ on the axis:
 \be
 r_s = r = sinh( \om t_s/\gamma_0){ v_0 \gamma_0  \over \om}
 \label{7}
 \ee
 where $\gamma_0 \equiv E_0=  1/\sqrt{1-v_0^2}$. 
 Thus, in terms of observer time $t_s$, motion of a particle 
  is nearly exactly the same as if we were to solve the  non-relativistic equation of motion $\ddot{r}= r \om^2$.
 Qualitatively, the reason is  that centrifugal force increases with $\gamma$, so that even though a particle becomes heavier, the centrifugal force increases proportionally. 
 In terms  of local observers  time, a particle starting from the axis with velocity $v_0$ reaches \LC\ in finite time $\Delta t_s = {\gamma_0 \over \om} arcsin(1/(v_0 \gamma_0))$, beyond which (\ref{7}) is inapplicable.

 It is somewhat surprising that an observer located infinitely close to the axis of rotation, and thus moving with infinitely small velocity with respect to the stationary observer on the axis, measures a qualitatively different acceleration (positive for rotating, negative for observer on the axis). This is due to the fact that the unit  frame vectors  describing the physical experience of  rotating  observers  are not Fermi-Walker transported along the world line; these   observers are spinning as well as non-inertial.

 \subsection{Radially falling particle in \Sc\  metric}
 
 As yet another way to look at this 
 controversial issue,   let us discuss  briefly a very similar problem with a known answer: radial falling of a particle into \Sc\ black hole.
In coordinate time \cite{Lightman}
 \ba &&
 (\partial_t r)^2 = (1- 2 M/r)^2 (1-(1-2M/r)E_0^2)
 \nn &&
 \partial_t^2 r = (1- 2 M/r){ M \over E_0^2 r^2 } \left({6 M \over  r} - 3 + 2 E_0^2 \right)
 \ea
 Thus, coordinate acceleration reverses at $ r= 6M/ ( 3 - 2 E_0^2)$. For a particle starting at rest at infinity this reversal occurs at $r=6M$. But for $E_0 > \sqrt{3/2}$, acceleration is always positive, directed away from the black hole.  Thus, if at infinity a particle is shot towards \BH\ with $\beta > 1/\sqrt{3}$ the cordinate accelrataion is always directed away from \BH. If we were to take this  mathematically correct result literally, it would mean that gravitational force becomes repulsive. 
 Of course, the resolution of the ``paradox'' in this case is obvious, and is similar to the ``reversal'' of acceleration in rotating frame: one cannot use coordinate acceleration to infer physically relevant quantities; one needs to use a defined set of observers. For any observer at fixed radius, 
 \ba &&
 (\partial_{t_s} r)^2 = (1- 2 M/r) (1-(1-2M/r)E_0^2)
 \nn &&
 \partial_{t_s} ^2 r = - {M \over E_0^2 r^2}
 \ea
 acceleration is always negative, towards a \BH\  (proper acceleration is  also negative $  \partial_\tau ^2 r = - {M / r^2}$).

 \subsection{Photon motion}
 
 Finally, let us show that  radial motion of a photon in rotating frame  (\eg\ along an optical fiber attached to the wire) experiences the same ``deceleration'' when measured in terms of coordinate time, as that of a relativistic particle. A condition $ds=0$ gives in rotating frame
 \be
 {dr_{\rm ph} \over dt} = \sqrt{1-  \om^2 r^2}
 \label{rphot}
 \ee
 This has formal solution $ r _{\rm ph} =(1/\om) \sin \om t$ for a photon emitted from $r=0$ at $t=0$. Surely, it does not mean that a photon bounces back from the \LC!
 Eq. (\ref{rphot}) measures coordinate velocity of a photon, which is not surprisingly differs from $c$. 
  
\section{Centrifugal effect in general relativity}

It is straightforward to repeat the previous derivations in  a coordinate  frame rotating in  \Sc\ metric.
Making  a coordinate transformation $\phi \rightarrow \phi' - \om t$ and
assuming  that in rotating  coordinates motion is strictly radial, $d\phi'=d\theta =0$, the non-vanishing components of the metric tensor are
\ba &&
g_{00} = - \left( 1 - {2 M \over r  } - \om^2 r^2 \right) 
\nn &&
g_{rr}= \left( 1 - {2 M \over r  } \right)^{-1}
\ea
 There are two {\LC}s, inner and outer, solutions of $1 - {2 M \over r }- \om^2 r^2 =0 $.   Since determinant of the metric tensor is $<0$ beyond the {\LC}s, approximation of a rigidly rotating wire is inapplicable in those regions (formally it becomes  applicable again inside the horizon).
  Inner and outer {\LC}s coinside when $\om=\om_{\rm ph}$, angular velocity of a photon orbit $r=3M$.
In case of \Sc\ \BH\, it is required that $\om < 1/(3 \sqrt{3} M)$.

 \subsection{Motion of a particle along the radial wire}
The Hamilton - Jacobi equation,
\be
{1\over  1 - {2 M \over r  } - \om^2 r^2} (\partial_t S)^2 - \left( 1 - {2 M \over r  } \right) (\partial_r S)^2 =1
\ee
gives
\ba &&
(\partial_t  r)^2 = \left( 1 - {2 M \over r  } \right)  \left( 1 - {2 M \over r }- \om^2 r^2  \right) 
\nn && \times
 \left( 1 - {1 - {2 M \over r }- \om^2 r^2 \over E_0^2}  \right)
 \label{vr}
 \ea
 For completeness we also give the relevant Christoffels 
 \ba &&
 \Gamma ^t_{tr}= \Gamma ^t_{rt}= {1\over 2 g_{tt} } \partial_r g_{tt} = { r\om^2 - M/r^2\over 1- 2M/r- r^2 \om^2}
\nn &&
\Gamma^r_{tt}= - {1\over 2 g_{rr} } \partial_r g_{tt} =(1-2 M/r)( r \om^2 -M/r^2)
\nn &&
\Gamma^r_{rr} ={1\over 2 g_{rr} } \partial_r g_{rr}=- M/(r(r- 2M))
\ea

Transforming to a local stationary observer rotating with the wire
   \ba&&
  dr_s ={dr  \over \sqrt{1 - {2 M \over r  }}}
 \nn &&
 dt_s = \sqrt{1 - {2 M \over r }- \om^2 r^2  } dt_s,
 \ea
 we find
\ba &&
(\partial_{t _s} r_s)^2 = 1- { 1 - {2 M \over r }- \om^2 r^2   \over E_0^2} 
\nn &&
 { \partial^ 2r_s \over \partial t_s^2 }= - 
{ \sqrt{1 - {2 M \over r }}  \over E_0} \left ( {M \over r^2 } - r \om^2 \right) =
\nn &&
- 
 \sqrt{1 - {2 M \over r }} {  (1 - (\partial_{t _s} r_s)^2)  \over 1 - {2 M \over r }- \om^2 r^2 }  \left[ {M \over r^2 } - r \om^2 \right]
 \label{AA1}
 \ea
  Again,  velocity (\ref{AA1}) is an invariant, as can be verified directly using  the four-velocity of a bead (which follows from (\ref{vr})) and $U_2=\{-1/\sqrt{ 1 - {2 M \over r }- \om^2 r^2},0\} $, the four-velocity of a stationary observer. The first term in square brakets can be identified with graviational acceleration, the second term - with \cfg\  acceleration.
 Inside the {\LC}s they  do not change signs.

  Finally, the equations of motion in  terms of a  proper time read
 \ba && 
\left({ \partial r \over \partial \tau}  \right)^2= \left( 1 - {2 M \over r }  \right) \left( { E_0^2 \over 1 - {2 M \over r }- \om^2 r^2} -1 \right)
\nn &&
{\partial^2 r \over \partial \tau^2} = - { M \over r^2} + { (r -3M)  \om^2 E_0^2 \over \left( 1 - {2 M \over r }- \om^2 r^2  \right)^2}= - { M \over r^2}  +(r -3M)  \om^2
\label{reverse}
\ea
where the last equality uses the fact that for circular orbit $E_0=-g_{00}= 1 - {2 M \over r }- \om^2 r^2$.
Eq. (\ref{reverse}) shows the reversal of centrifugal force at the photon circular orbit  $r= 3M$ \cite{AbramowiczPrasanna}.
Thus, the proper observer sees a reversal at $r= 3M$.

For Kerr \BH,  no clear separation can be made between the effects of the wire rotation and rotation of space-time, so that  a notion of a centrifugal force becomes not well defined, see  \cite{AbramowiczPrasanna,deFelice91,AbramowiczLasota97,IyerPrasanna} for discussion.
\section{Discussion}

We have discussed a subtle special relativistic effect, the seeming reversal of centrifugal acceleration for relativistically moving particle. Straightforward analysis seems to indicate that centrifugal acceleration reverses its direction for fast moving particles, and becomes centrifugal deceleration, which seems to prevent a particle escaping from the system. This conclusion was drawn by  Machabeli \& Rogava    \cite{MachabeliRogava} and applied to a number of astrophysical cases. It is mathmatically correct, but physical interpretation that   centrifugal acceleration reverses in rotating frame is wrong, since the
motion was defined in a frame-dependent  way. It is the coordinate acceleration which reverses, while any physically relevant acceleration, \eg\ measured by a set of stationary  observers and/or proper acceleration remain directed away from the axis of rotation.   As a result, a change in the velocity Machabeli \& Rogava      \cite{MachabeliRogava} found reflects mostly the changing rate of time measured by locally stationary observers and not the motion of a bead.
This is similar to freezing of motion on the horizon of a \BH\ for a free-falling particle, when considered in \Sc\ coordinates.

The centrifugal acceleration  controversy provides an excellent illustration of 
one of the principal issues in GR,  that physical effects should be formulated in a frame-independent, but observer-dependent form (\eg\ a set of ZAMO observers). For  a defined set of  observers, \eg\ stationary with respect to the wire,  a particle 
always accelerates and 
reaches the speed of light when crossing the \LC. 
The analogy between bead on a wire and free fall motion in \Sc\ geometry is nearly exact: in both cases a particle reaches a speed of light while approaching the point where $g_{00}=0$,  \LC\ or horizon. The only difference is that in case of a rotating wire beyond the \LC\ the determinant of the metric tensor becomes negative, so that the system becomes unphysical, while the determinant of the  metric tensor remains positive when crossing the horizon.

I would like to thank Roger Blandford, Ilya Mandel, Saul Teukolsky and Sergey Khlebnikov for discussions.

\bibliographystyle{apsrev}
\bibliography{/Users/maximlyutikov/Home/Research/BibTex}

\appendix

\end{document}